\documentclass[twocolumn,showpacs,aps,prl]{revtex4}
\usepackage{graphicx}

\begin{document}

\title{
Phase Diagram of One-Dimensional Extended Hubbard Model at Half Filling
}

\author{M. Tsuchiizu}
\author{A. Furusaki}
\affiliation{Yukawa Institute for Theoretical Physics, 
         Kyoto University, Kyoto 606-8502, Japan}
\date{\today}

\begin{abstract} 
We reexamine the ground-state phase diagram
   of the one-dimensional half-filled
   Hubbard model with on-site and 
   nearest-neighbor repulsive interactions. 
We calculate second-order corrections to coupling constants
   in the $g$-ology to show that the bond-charge-density-wave (BCDW)
   phase exists for weak couplings in between
   the charge density wave (CDW) and spin density wave (SDW) phases.
We find that the umklapp scattering of parallel-spin electrons
   destabilizes the BCDW state and gives rise to a bicritical point
   where the CDW-BCDW and SDW-BCDW continuous-transition lines merge
   into the CDW-SDW first-order transition line.
\end{abstract}


\maketitle

Electronic correlations in solids
   have been a subject of intensive research over the years.
Correlation effects have the strongest impact at commensurate band
   filling, where a system often undergoes a Mott transition.
The one-dimensional (1D) extended Hubbard model (EHM) with the
   nearest-neighbor repulsion $V$,
   in addition to the on-site repulsion $U$, is a simple, but
   nontrivial model that exhibits rich phase structure \cite{Emery}.
The model has a long history of research, and considerable amount of
   knowledge has been accumulated.
Much effort has been devoted to understanding its ground-state
   phase diagram at half filling.
In the strong-coupling limit\cite{Emery,Bari,Hirsch,Dongen} one can
   show that the model has two insulating phases, the
   spin-density-wave (SDW) phase and the charge-density-wave (CDW)
   phase, which are separated by a first-order transition line located 
   at $U\simeq2V$.
In the weak-coupling limit the perturbative renormalization group (RG)
   analysis\cite{Emery} concluded that there is a continuous phase
   transition between the CDW and SDW phases also at $U=2V$.
It was then considered that, as the coupling constants increase, the
   continuous-transition line changes into the first-order one at a
   tricritical point in the intermediate-coupling regime.
This picture was supported by both
   numerical\cite{Hirsch,Cannon,Zhang} and
   analytical\cite{Cannon,Voit} studies and had been regarded
   as the complete phase diagram of the EHM at half filling.

Quite recently, however, Nakamura\cite{Nakamura} found numerically
   that another phase exists between the CDW and SDW phases for weak
   couplings.
The new phase is the bond-charge-density-wave (BCDW) phase in which
   the Peierls dimerization occurs spontaneously.
He concluded that SDW-BCDW and BCDW-CDW transitions are continuous and
   that these two transition lines merge at a multicritical point into
   the first-order line separating the CDW and SDW phases \cite{note}.
His claim was confirmed by a recent extensive Monte Carlo
   calculation \cite{Sengupta}.
The appearance of a spontaneously dimerized phase in the EHM is
   surprising and calls for thorough theoretical study.
So far the BCDW phase has been shown analytically to exist
   only in models with extra correlated-hopping interactions
   \cite{Japaridze}.
For the original EHM, however, the origin of the BCDW phase and the
   nature of the associated phase transitions are not fully understood.
In this Letter, we will provide theoretical argument for the existence
   of the BCDW phase by reformulating the weak-coupling theory to
   include higher-order terms.
Using the bosonization technique, we derive a set of RG equations and
   discuss the critical properties of the phase transitions.
We find that the umklapp scattering between electrons with parallel
   spins is responsible for the emergence of the bicritical point.

The Hamiltonian of the 1D EHM is
\begin{eqnarray}
H &=&
 - t \sum_{j,\sigma}
   \left( c_{j,\sigma}^\dagger c_{j+1,\sigma} + \mathrm{h.c.}\right)
\nonumber \\
&&{}
 + U \sum_j n_{j,\uparrow} \, n_{j,\downarrow}
   + V \sum_j n_{j} \, n_{j+1} ,
\label{eq:H1D}
\end{eqnarray}
  where
  $n_{j,\sigma} \equiv c_{j,\sigma}^\dagger c_{j,\sigma}-\frac{1}{2}$,
  $n_j \equiv n_{j,\uparrow} + n_{j,\downarrow}$, and
  $c_{j,\sigma}^\dagger$ denotes the creation operator of an
  electron at the $j$th site with spin $\sigma$.
Following the previous studies on models with correlated-hopping
  interactions\cite{Japaridze}, we consider the CDW, SDW, BCDW and 
  bond-spin-density-wave (BSDW) phases.
They are characterized by the order parameters,
  $\mathcal{O}_\mathrm{CDW}\equiv(-1)^j n_j$,
  $\mathcal{O}_\mathrm{SDW}\equiv(-1)^j(n_{j,\uparrow}-n_{j,\downarrow})$,
  $\mathcal{O}_\mathrm{BCDW}\equiv(-1)^j\sum_\sigma
     (c^\dagger_{j,\sigma}c_{j+1,\sigma}^{}+\mathrm{h.c.})$,
  and
  $\mathcal{O}_\mathrm{BSDW}\equiv(-1)^j
     (c^\dagger_{j,\uparrow}c_{j+1,\uparrow}^{}
      -c^\dagger_{j,\downarrow}c_{j+1,\downarrow}^{}
      +\mathrm{h.c.})$
  \cite{Nersesyan}.

\begin{figure}
\includegraphics[width=8cm]{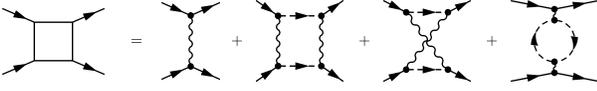}
\caption{ 
Vertex diagrams.
Solid lines denote the low-energy states and the dashed lines
represent the high-energy states to be integrated out.
}
\label{fig:diagram} 
\end{figure}

We first focus on the weak-coupling limit $U,V\ll t$.
The hopping $t$ generates the energy band with dispersion
$\varepsilon_k=-2t\cos k$.
At half filling the Fermi points are at $k=\pm k_F=\pm\pi/2a$,
where $a$ is a lattice constant.
Electrons experience two-particle scattering by the on-site and
nearest-neighbor repulsions $U$ and $V$. 
We follow the standard $g$-ology approach \cite{Emery,Solyom} and
parametrize the scattering matrix elements by the coupling constants
$g$.
In lowest order in $U$ and $V$ they are known to be
$g_{1\parallel}=-2Va$, $g_{1\perp}=(U-2V)a$, $g_{2\parallel}=2Va$,
$g_{2\perp}=(U+2V)a$, $g_{3\parallel}=-2Va$, $g_{3\perp}=(U-2V)a$,
$g_{4\parallel}=2Va$, $g_{4\perp}=(U+2V)a$, where we have used the
standard notation \cite{Emery,Solyom}.
Here we note that both $g_{1\perp}$ and $g_{3\perp}$ vanish at $U=2V$, 
and this is the reason why the lowest-order calculation predicts the
direct CDW-SDW transition at $U=2V$.
Hence, we need to go beyond the lowest order to see if the
BCDW phase really exists.
To this end, we adapt the two-step RG scheme used in
Ref.~\cite{Penc_Mila}.
(i) We separate the states into low-energy states ($||k|-k_F|<\Lambda$)
and high-energy ones ($||k|-k_F|>\Lambda$) by introducing a momentum
cutoff $\Lambda$, and integrate out high-energy states to obtain
effective scattering matrix elements for low-energy states.
(ii) We then derive one-loop RG equations for these matrix elements
using the standard bosonization method.
The diagrams for the effective couplings up to second order in $U$
and $V$ are shown in Fig.~\ref{fig:diagram}.
The explicit calculation yields
\begin{eqnarray}
g_{1\perp} &=&
(U-2V)a \left[1
- \frac{C_1}{4\pi t}(U-2V)\right]
-\frac{C_2}{\pi t} \, V^2a
,
\label{eq:g1perp}
\\
g_{3\perp} &=& 
(U-2V)a\left[1
+\frac{C_1}{4\pi t}(U+6V)\right]
+\frac{C_2}{\pi t} \, V^2a
,
\label{eq:g3perp}
\end{eqnarray} 
   where
   $C_1(\Lambda)\equiv 2\ln [\cot (a\Lambda/2)]>0$ and
   $C_2(\Lambda)\equiv 2\cos(a \Lambda) >0$.
The weak dependences of $C_i$s on $\Lambda$ allow us to set
   $\Lambda=\pi/4a$; different choices will only lead to small
   quantitative changes.
We see that $g_{1\perp}<0$ and $g_{3\perp}>0$ at $U=2V$ due to
   the $C_2$ term.
This implies that a new phase different from the CDW and SDW can
   appear for $U\simeq 2V$, as we will show shortly.
The zeros of $g_{1\perp}$ and $g_{3\perp}$ are shifted
   from $U=2V$ due to the  momentum dependence of the
   matrix element $2Va\cos(qa)$ for the virtual scattering of
   high-energy states ($q\ne 0,2k_F$).
There is no symmetry principle that enforces $g_{1\perp}$ and
   $g_{3\perp}$ to vanish simultaneously.

Having derived the effective scattering matrix elements for low-energy 
states, we now apply the bosonization method.
The right-going and left-going electron fields $\psi_{\pm,\sigma}$ are 
written \cite{Emery,Solyom} 
\begin{eqnarray}
\psi_{p,\sigma}(x)
  &=&
   \frac{\eta_\sigma}{\sqrt{2\pi a}}
   \exp\left[ipk_F x+ip \, \varphi_{p,\sigma}(x)\right],
\label{eq:field_op}
\end{eqnarray}
where $\varphi_{p,\sigma}$ $(p=+/-)$ are the chiral bosonic fields
   and $\{\eta_\sigma,\eta_{\sigma'}\}=2\delta_{\sigma,\sigma'}$.
The bosonic fields obey the commutation relations
   $\left[ \varphi_{p,\sigma}(x),\varphi_{p,\sigma'}(x') \right]
    = ip \pi \,\mathrm{sgn}(x-x') \, \delta_{\sigma,\sigma'}$ and
   $\left[ \varphi_{+,\sigma}(x),\varphi_{-,\sigma'}(x') \right]
    = i\pi \delta_{\sigma,\sigma'}$.
We define chiral charge fields,
   $\theta_p=(\varphi_{p,\uparrow}+\varphi_{p,\downarrow})/2$, and
   chiral spin fields,
   $\phi_p=(\varphi_{p,\uparrow}-\varphi_{p,\downarrow})/2$,
   to write the Hamiltonian density for low-energy states:
\begin{eqnarray}
\mathcal{H} &=&
\frac{1}{2\pi}\sum_{p=+,-}
 \left[v_\rho(\partial_x\theta_p)^2
      +v_\sigma(\partial_x\phi_p)^2\right]
\nonumber \\ && {}
+ \frac{g_\rho}{2\pi^2}
     \left(\partial_x \theta_+ \right)
     \left(\partial_x \theta_- \right)
- \frac{g_\sigma}{2\pi^2}
     \left(\partial_x \phi_+ \right) 
     \left(\partial_x \phi_- \right)
\nonumber \\ && {}
-\frac{g_c}{2(\pi a)^2} \, \cos 2 \theta 
+\frac{g_s}{2(\pi a)^2} \, \cos 2 \phi
\nonumber \\ && {}
-\frac{g_{cs}}{2(\pi a)^2} \, \cos 2\theta  \,  \cos 2\phi
\nonumber \\ && {}
-\frac{g_{\rho s}}{2 \pi^2}
     \left(\partial_x \theta_+ \right) 
     \left(\partial_x \theta_- \right) \, \cos 2\phi
\nonumber \\ && {}
+\frac{g_{c\sigma}}{2 \pi^2}
     \left(\partial_x \phi_+\right) 
     \left(\partial_x \phi_- \right) \, \cos 2\theta 
\nonumber \\ && {}
+\frac{g_{\rho \sigma}}{2\pi^2}  \, a^2 
     \left(\partial_x \theta_+ \right) 
     \left(\partial_x \theta_- \right) 
     \left(\partial_x \phi_+ \right) 
     \left(\partial_x \phi_- \right) 
,
\label{eq:Hamiltonian}
\end{eqnarray}
   where $\theta=\theta_+ + \theta_-$, $\phi=\phi_+ + \phi_-$.
The renormalized velocities are
   $v_\rho=2ta+(g_{4\parallel}+g_{4\perp}-g_{1\parallel})/2\pi$ and
   $v_\sigma=2ta+(g_{4\parallel}-g_{4\perp}-g_{1\parallel})/2\pi$.
To simplify the notation, we have written $g_c=g_{3\perp}$,
   $g_s=g_{1\perp}$, and $g_{cs}=g_{3\parallel}$.
The other coupling constants are given by
   $g_\rho=g_{2\perp}+g_{2\parallel}-g_{1\parallel}$,
   $g_\sigma=g_{2\perp}-g_{2\parallel}+g_{1\parallel}$, and
   $g_{\rho s}=g_{c\sigma}=g_{\rho\sigma}=-2Va$ to lowest order
   in $V$.
The $g_{\rho s}$ ($g_{\rho\sigma}$) coupling comes from the backward
   scattering of electrons with opposite (parallel) spins, while the
   $g_{c\sigma}$ coupling is generated
   from the umklapp scattering of electrons with antiparallel spins.
The SU(2) symmetry in the spin sector ensures $g_\sigma=g_s$,
   $g_{cs}=g_{c\sigma}$, and $g_{\rho s}=g_{\rho\sigma}$,
   and therefore it is important to retain the $g_{\rho\sigma}$ term.

\begin{figure}
\includegraphics[width=7cm]{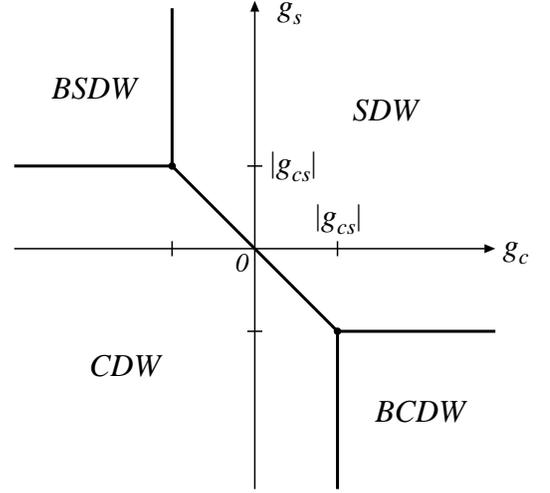}
\caption{
Phase diagram obtained by minimizing the potential energy.
Bicritical points are at $g_c=-g_s=\pm g_{cs}$.
}
\label{fig:classic}
\end{figure}

In terms of the phase fields $\theta$ and $\phi$
 the order parameters are written as
\begin{eqnarray}
\mathcal{O}_\mathrm{SDW}(x)
&\propto&
\cos\theta(x) \, \sin \phi(x),
\\
\mathcal{O}_\mathrm{CDW}(x)
&\propto&
\sin\theta(x) \, \cos \phi(x),
\\
\mathcal{O}_\mathrm{BCDW}(x)
&\propto&
\cos\theta(x) \, \cos \phi(x),
\\
\mathcal{O}_\mathrm{BSDW}(x)
&\propto&
\sin\theta(x) \, \sin \phi(x).
\end{eqnarray}
The phase diagram can be qualitatively understood via a
   quasi-classical analysis: we neglect spatial variations
   of the fields and focus on the potential,
   $V(\theta,\phi)=
    - g_c\cos2\theta + g_s\cos 2\phi
    - g_{cs} \cos2\theta \, \cos 2\phi$,
   where $g_{cs}=g_{3\parallel}<0$.
The order parameters take maximum amplitudes when the fields
   $\theta$ and $\phi$ are pinned at the following
   potential minima:
   $(\theta,\phi)=(0,\pm\pi/2)$ in the SDW state,
   $(\pm\pi/2,0)$ in the CDW state,
   $(0,0)$ or $(\pi,0)$ in the BCDW state, and
   $(\pi/2,\pm\pi/2)$ in the BSDW state (mod $\pi$).
In these states the potential energy $V(\theta,\phi)$ becomes
   $V_\mathrm{SDW} 
                =-g_c-g_s-|g_{cs}|$,  
   $V_\mathrm{CDW} 
                =g_c+g_s-|g_{cs}|$,
   $V_\mathrm{BCDW} 
                 =-g_c+g_s+|g_{cs}|$ and
   $V_\mathrm{BSDW} 
                 =g_c-g_s+|g_{cs}|$,
   respectively.
Comparing these energies, we obtain the phase diagram in the
   $g_c$-$g_s$ plane (Fig.~\ref{fig:classic}).
The direct CDW-SDW transition is first order
   because there is a potential barrier of height
   $\min(|g_{cs}|,2|g_{cs}| - 2|g_c|)$
   between the corresponding minima.
The other boundaries located at $g_s=\pm|g_{cs}|$ and
   $g_c=\pm|g_{cs}|$ are continuous transitions, because the pinning
   potential for $\theta$ or $\phi$ vanishes when the other phase field
   is pinned.
When $g_{cs}=0$, the first-order CDW-SDW transition line collapses to
   a tetracritical point.

To obtain the ground-state phase diagram of the EHM, we need to
   include the renormalization of the coupling constants due to
   quantum fluctuations of the fields.
A systematic analysis in the weak-coupling limit can be done by
   applying the perturbative RG method to $\mathcal{H}$
   (\ref{eq:Hamiltonian}).
The SU(2) spin symmetry guarantees the relations $g_\sigma=g_s$,
   $g_{c\sigma}=g_{cs}$, and $g_{\rho\sigma}=g_{\rho s}$ to hold
   in the scaling procedure.
The one-loop RG equations that describe changes of the coupling
   constants during the scaling of the short-distance cutoff
   ($a\to ae^{dl}$) are then given by
\begin{eqnarray}
\frac{d}{dl} G_\rho 
  &=& {}
    + 2 \, G_c^2 + G_{cs}^2 +  G_s \, G_{\rho s} ,
\label{eq:Grho}
\\
\frac{d}{dl} G_c 
  &=& {}
    + 2 \, G_\rho\, G_c - G_s \, G_{cs} - G_{cs} \, G_{\rho s} ,
\label{eq:Gc}
\\
\frac{d}{dl} G_s 
  &=& {}
    - 2 \, G_s^2 - G_c \, G_{cs} - G_{cs}^2 ,
\label{eq:Gs}
\\
\frac{d}{dl} G_{cs} 
  &=& {}
    - 2 \, G_{cs} + 2 \, G_\rho \, G_{cs} - 4 \, G_s \, G_{cs}
\nonumber \\ && {}
    - 2 \, G_c\, G_{s}
    - 2 \, G_c\, G_{\rho s}
    - 4 \, G_{cs} \, G_{\rho s} ,
\label{eq:Gcs}
\\
\frac{d}{dl} G_{\rho s} 
  &=& {}
    - 2 \, G_{\rho s} + 2 \, G_\rho \, G_s
\nonumber \\ && {}
    - 4 \, G_c \, G_{cs} - 4 \, G_{cs}^2
    - 4 \, G_s \, G_{\rho s} ,
\label{eq:Grhos}
\end{eqnarray}
   where $G_\nu=g_\nu/(4\pi ta)$.
From Eqs.~(\ref{eq:Gcs}) and (\ref{eq:Grhos}) one finds that $g_{cs}$
   and $g_{\rho s}$ are irrelevant and renormalized towards zero
   for weak interactions.
It is therefore natural to ignore $g_{cs}$ and $g_{\rho s}$ first.
With this approximation the Hamiltonian reduces to two decoupled
   sine-Gordon models, and it is easy to follow the RG flows of
   $G_\rho$, $G_c$, and $G_s$
   from Eqs.~(\ref{eq:Grho})--(\ref{eq:Gs}).
Since $g_\rho=(U+6V)a >0$, $G_c$ is relevant and grows at low
   energies.
The coupling $G_s$ is marginally
   relevant (irrelevant) for $g_s<0$ ($g_s>0$).
The phase diagram of the EHM is obtained tentatively from
   Fig.~\ref{fig:classic}
   by setting $g_c=g_{3\perp}$ [Eq.~(\ref{eq:g3perp})],
   $g_s=g_{1\perp}$ [Eq.~(\ref{eq:g1perp})], and $g_{cs}=0$.
When $U$ is sufficiently larger than $2V$ such that $g_c>0$ and
   $g_s>0$, we have the SDW phase.
If $U$ is smaller than $2V$ ($g_c<0$ and $g_s<0$),
   then we have the CDW phase.
Around the $U=2V$ line we find the BCDW phase, where
   $g_{1\perp}<0$ and $g_{3\perp}>0$ due to the $C_2$ term.
The BSDW phase does not exist in the EHM.
The charge excitations are gapful except on the CDW-BCDW transition
   line where the relevant pinning potential vanishes.
The spin excitations are gapless in the SDW phase and on the
   SDW-BCDW transition line and gapful otherwise.
In the gapped phases the charge gap $\Delta_c$ and the spin gap
   $\Delta_s$ are given by
   $\Delta_c\simeq t |G_c|^{1/2G_\rho}$ and
   $\Delta_s\simeq  t\exp(1/2G_s)$ for $|G_c|\ll1$ and
   $0<-G_s\ll1$, respectively.

Next we examine effects of the parallel-spin umklapp scattering
   $g_{cs}$ for $U\simeq2V$.
Let us assume $U-2V = -C_2V^2/\pi t + \mathcal{O}(V^3/t^2)$, i.e.,
   $g_c\approx0$ and $g_s<0$.
We are considering the situation very close to the CDW-BCDW
   transition.
In this case the spin gap is formed first as the energy scale is
   lowered, and we can replace $\cos2\phi$ with its average
   $\langle\cos2\phi\rangle\simeq(\Delta_s/t)^2$ for energies
   below the spin gap.
This means that the $\cos2\theta$ potential that tries to pin the
   fluctuating $\theta$ field has the effective coupling
\begin{eqnarray}
g_{c}^* &=& g_c + g_{cs} \langle\cos2\phi\rangle.
\label{eq:gc*}
\end{eqnarray}
The CDW-BCDW transition occurs when $g_c^*=0$, i.e.,
   $g_c=-g_{cs}\langle\cos2\phi\rangle >0$.
The phase space of the BCDW state is reduced upon inclusion of the
   $g_{cs}$ term.
Note, however, that the CDW-BCDW boundary does not move across
   the $U=2V$ line because
   $|g_{cs}\langle\cos2\phi\rangle|\simeq2Va\exp[-c(t/V)^2]$
   is much smaller than the $C_2$ term for $V\ll t$,
   where $c$ is a positive constant.
A similar argument applies to the region near the SDW-BCDW
   transition.
Suppose that $U-2V=+C_2V^2/\pi t + \mathcal{O}(V^3/t^2)$
   ($g_s\approx0$ and $g_c>0$).
In this case, as the energy scale is lowered, the charge gap
   opens first and the $\theta$ field is pinned at $\theta=0$
   (mod $\pi$).
Below the charge-gap energy scale the $\phi$ field is subject to
   the pinning potential $g_s^*\cos2\phi$ with
\begin{eqnarray}
g_{s}^* &=& g_s - g_{cs} \langle\cos2\theta\rangle,
\label{eq:gs*}
\end{eqnarray}
where $\langle\cos2\theta\rangle\simeq(\Delta_c/t)^{2(1-G_\rho)}$.
The SDW-BCDW transition now happens at
   $g_s=g_{cs}\langle\cos2\phi\rangle<0$,
   and thus the SDW-BCDW transition line moves to increase the
   SDW phase.
Again the phase boundary is not changed beyond the $U=2V$ line as
   $|g_{cs}\langle\cos2\theta\rangle|\simeq2Va(c'V/t)^{\pi t/V}$
   is much smaller than $V^2a/t$, where $c'$ is a constant of order
   1. 
This completes our proof of the existence of the BCDW phase near
   the $U=2V$ line in the weak-coupling limit.

For larger $U$ and $V$, the $g_{cs}$ coupling becomes less
   irrelevant, and the BCDW phase will eventually disappear.
Since the cosine factor in Eqs.~(\ref{eq:gc*}) and (\ref{eq:gs*})
   can be considered as renormalization of $g_{cs}$,
   we conclude that the two continuous lines meet
   when the renormalized couplings satisfy the relation
\begin{equation}
G_c=-G_s=-G_{cs}\equiv G \qquad (G>0)
\label{eq:condition}
\end{equation}
   in the low-energy limit.
This is the condition for the bicritical point in the RG scheme.
Note that the condition is not simply that the $g_{cs}$ term becomes
   relevant, as previously assumed \cite{Cannon}.
When Eq.~(\ref{eq:condition}) is satisfied, the effective potential
   takes a simple form
   $V(\theta,\phi)=-G(\cos2\theta+\cos2\phi-\cos2\theta\cos2\phi)$,
   which has an interesting feature that its potential minima are
   not isolated points but the crossing lines $\theta=\pi m$ or
   $\phi=\pi n$ ($m$, $n$: integer).
On these lines either $\theta$ or $\phi$ becomes a free field;
   the theory has more freedom than a single free bosonic field, but
   less than two free bosonic fields.
We thus expect that the theory of the bicritical point should have a
   central charge larger than 1 but smaller than 2.
Detailed analysis of the critical theory is left for a future study.

\begin{figure}
\includegraphics[width=7cm]{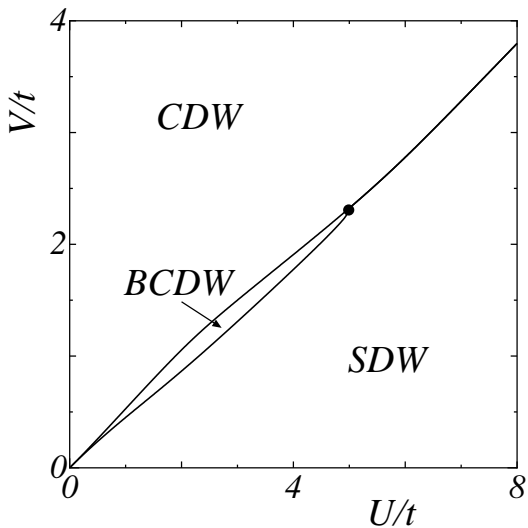}
\caption{
Phase digram of the half-filled extended Hubbard model.
The bicritical point is at $(U_c,V_c)\simeq(5.0t, 2.3t)$.
}
\label{fig:phase2}
\end{figure}

We have numerically solved the scaling equations
   (\ref{eq:Grho})-(\ref{eq:Grhos})
   to obtain the global phase diagram of the EHM.
The phase is determined by looking at which of the couplings
   $G_c$, $G_s$, and $G_{cs}$ becomes relevant.
The idea is essentially the same as what we have discussed above.
If $|G_c|$ grows with increasing $l$ and reaches, say, 1 first among
   the 3 couplings, then
   we stop the integration and calculate $G_s^*=G_s-G_{cs}$.
Since the charge fluctuations are suppressed below this energy scale,
   we are left with Eq.~(\ref{eq:Gs}) where $G_s$ replaced by $G_s^*$
   and $G_{cs}=0$.
We immediately see that a positive (negative) $G_s^*$ leads to the
   SDW (BCDW) state.
If $|G_s|$ becomes 1 first, then the sign of $G_c^*=G_c-G_{cs}$
   determines the phase: the CDW (BCDW) state for $G_c^*<0$
   ($G_c^*>0$).
Finally, when $|G_{cs}|$ reaches 1 first, we stop the calculation and
   compare $G_c$ and $G_s$.
Since both charge and spin fluctuations are already suppressed by the
   $G_{cs}\cos2\theta\cos2\phi$ potential, we can deduce the phase
   from the quasi-classical argument.
From Fig.~\ref{fig:classic} we see that we have the SDW state for
   $G_s>-G_c$ and the CDW state for $G_s<-G_c$.
In the SDW state the pinning potential for the $\phi$ field is
   marginally irrelevant, and therefore the spin sector becomes gapless.
The phase diagram obtained in this way is shown in Fig.~\ref{fig:phase2}.
For weak couplings the BCDW phase appears at $U\simeq 2V$, and
   the successive continuous transitions between the SDW, BCDW and
   CDW states occur as $V/U$ increases.
As $U$ and $V$ increase along the line $U=2V$, the BCDW phase first
   expands and then shrinks up to the bicritical point
   $(U_c,V_c)\approx(5.0t, 2.3t)$ where the two continuous-transition
   lines meet.
Beyond this point the BCDW phase disappears and we have the direct
   first-order transition between the CDW and the SDW phases.
The phase diagram (Fig.~\ref{fig:phase2}) is similar to the one
    reported recently \cite{Nakamura,Sengupta}.
The position of the phase boundaries in Fig.~\ref{fig:phase2} is
   not reliable quantitatively however, as we have used the
   perturbative RG equations which are valid only in the
   weak-coupling regime.

In summary, we have studied the ground-state phase diagram of
   the 1D extended Hubbard model with repulsive 
   interaction at half filling.
We have shown analytically that the BCDW phase appears
   at $U\simeq 2V$ in the weak-interaction limit.
We have also discussed the instability of the BCDW state and
   the emergence of the bicritical point due to the parallel-spin
   umklapp scattering.

M.T.~thanks Y.~Suzumura, H.~Yoshioka, M.~Sugiura, and,
   especially, E.~Orignac for valuable discussions.
This work was supported in part by Grant-in-Aid for
   Scientific Research on Priority Areas (A) from The Ministry of
   Education, Science, Sports and Culture (No.~12046238) and by
   Grant-in-Aid for Scientific Research (C) from Japan Society for the
   Promotion of Science (No.~10640341).


\begin{thebibliography}{}
\bibitem{note}
Since the topology of the phase diagram is changed,
the term ``tricritical'' is no longer appropriate for the
multicritical point.
Instead we will call it the bicritical point.
\end{thebibliography}
\end{document}